\documentclass[preprints,article,accept,moreauthors,pdftex]{mdpi}

\graphicspath{ {./images/} }

\defcitealias{soldateschi_2020}{SBD20}
\defcitealias{soldateschi_2021}{SBD21}

\firstpage{1}
\pubvolume{1}
\issuenum{1}
\articlenumber{0}
\pubyear{2021}
\copyrightyear{2020}
\datereceived{}
\dateaccepted{}
\datepublished{}
\hreflink{https://doi.org/} % If needed use \linebreak
\pdfoutput=1

\Title{Detectability of continuous gravitational waves from magnetically deformed neutron stars}

\TitleCitation{Detectability of continuous gravitational waves from magnetically deformed neutron stars}

\Author{Jacopo Soldateschi $^{1,2,3}$*\orcidA{} and Niccolò Bucciantini $^{2,1,3}\orcidB{}$}

\AuthorNames{Jacopo Soldateschi and Niccolò Bucciantini}

\AuthorCitation{Soldateschi, J.; Bucciantini, N.}
\address{%
$^{1}$ \quad Dipartimento di Fisica e Astronomia, Universit\`a degli Studi di Firenze, Via G. Sansone 1, I-50019 Sesto F. no (Firenze), Italy\\
$^{2}$ \quad INAF - Osservatorio Astrofisico di Arcetri, Largo E. Fermi 5, I-50125 Firenze, Italy\\
$^{3}$ \quad INFN - Sezione di Firenze, Via G. Sansone 1, I-50019 Sesto F. no (Firenze), Italy
}

\corres{Correspondence: jacopo.soldateschi@inaf.it}

\abstract{Extremely powerful magnetic fields are contained inside neutron stars. Their effect is to deform the shape of the star, leading to the emission of continuous gravitational waves. The magnetic deformation of neutron stars depends on the details of their magnetic field, that is its geometry and strength. Moreover, it depends on their composition, described by the equation of state. Unfortunately, both the configuration of the magnetic field and the equation of state of neutron stars are unkown, and assessing the detectability of continuous gravitational waves from neutron stars suffers from these uncertainties. Using our recent results relating the magnetic deformation of a neutron star to its mass and radius, and considering the Galactic pulsar population, we assess the detectability of continuous gravitational waves from pulsars in the Galaxy - described by realistic equations of state currently allowed by observational and nuclear physics constraints - by gravitational waves detectors.}


\begin{document}
%%%%%%%%%%%%%%%%%%%%%%%%%%%%%%%%%%%%%%%%%%
% \setcounter{section}{-1} %% Remove this when starting to work on the template.
% \section{How to Use this Template}
%
% The template details the sections that can be used in a manuscript. Note that the order and names of article sections may differ from the requirements of the journal (e.g., the positioning of the Materials and Methods section). Please check the instructions on the authors' page of the journal to verify the correct order and names. For any questions, please contact the editorial office of the journal or support@mdpi.com. For LaTeX-related questions please contact latex@mdpi.com.
%The order of the section titles is: Introduction, Materials and Methods, Results, Discussion, Conclusions for these journals: aerospace,algorithms,antibodies,antioxidants,atmosphere,axioms,biomedicines,carbon,crystals,designs,diagnostics,environments,fermentation,fluids,forests,fractalfract,informatics,information,inventions,jfmk,jrfm,lubricants,neonatalscreening,neuroglia,particles,pharmaceutics,polymers,processes,technologies,viruses,vision

\section{Introduction}
The most dense material objects in the known Universe are neutron stars (NSs). While the hypothesis of their existence dates back to the 1930s \citep{landau_1932,baade_1934}, their actual discovery happened more than thirty years later.
In 1967 it was pointed out that if NSs were spinning and harboured strong magnetic fields, they would emit electromagnetic waves \citep{pacini_1967}; during the same year, regular radio pulses coming from a region of the sky were discovered \citep{hewish_1968}, and this `pulsar' was later interpreted to be a NS \citep{gold_1968}. Since then, thousands of NSs were discovered \citep{atnf_2005}, most of them as pulsars. A special kind of pulsar are millisecond pulsars (MSPs), pulsars with rotation periods under $\sim 20$ms.
\\
Another class of NSs is that of magnetars, possessing an extremely strong magnetic field, among the most powerful ever detected \citep{duncan_1992,thompson_1993,thompson_1995,thompson_1996}. While initially discovered as different high energy sources showing either energetic bursting (soft gamma repeaters) or periodic variability (anoumalous X-ray pulsars)  \citep{kouveliotou_1998,gavriil_2002,mereghetti_2015}, they were later shown to be part of the same class of objects. Even though the observed magnetar population is tiny (just over 30 sources \citep{olausen_2014}) when compared to the known NS population, it is believed that they might actually compose a significant fraction of the young population \citep{kaspi_2017}. While the magnetic field at the surface of pulsars has been inferred to be in the range $10^{8-12}$G \citep{asseo_2002,spruit_source_2009,ferrario_magnetic_2015}, in  magnetars it is thought to be able to reach $10^{15}$G \citep{olausen_2014,popov_origins_2016} and even $10^{17-18}$G in the case of newly-born proto-NSs \citep{del_zanna_chiral_2018,ciolfi_2019,franceschetti_2020}.
\\
Unfortunately, the geometry and strength of their internal magnetic fields remains mostly unconstrained: what is known is that neither purely poloidal nor purely toroidal configurations are stable \citep{prendergast_equilibrium_1956,chandrasekhar_1956,chandrasekhar_1957,tayler_adiabatic_1973,markey_1973,markey_1974,tayler_1980}, thus favouring mixed field configurations like the twisted torus \citep{ciolfi_2013,uryu_equilibrium_2014,pili_axisymmetric_2014}. In any case, magnetic fields of such strength cause potentially observable variations in the phenomenology of NSs, like a modification of their torsional oscillations \citep{samuelsson_2007,sotani_2015}, of their cooling properties \citep{page_2004,aguilera_2008} and a deformation in their shape \citep{haskell_2008,gomes_2019}.
\\
Also the internal structure of matter in NSs, encoded by the equation of state (EoS), remains largely unknown, enriching and complicating the scenario, even if the observation of NSs with a mass higher than $2$M$_\odot$ \citep{kandel_2020,Fonseca_2021} has partly constrained the problem, ruling out many EoSs. Meanwhile, results of the NICER telescope \citep{pang_2021,zhang_2021} have set tighter constraints also on the possible radii, effectively shrinking the allowed region of the mass-radius diagram of NSs and thus further constraining their EoS. In addition, the first observation of gravitational waves (GWs) emitted by a binary NS merger \citep{abbott_gw170817:_2017} allowed us to set limits on the stiffness of the EoS \citep{abbott_2018_1}. While the internal magnetic field and the EoS of NSs are the two major unknowns in their physics, they are deeply intertwined: strong magnetic fields directly affect the particle composition of NSs, playing a role in particle physics issues like the Delta puzzle \citep{cai_2015,Drago_Lavagno_Pagliara_Pigato_2016}, the hyperon puzzle \citep{zdunik_2013,chatterjee_2015}, the hadron-quark phase transition \citep{avancini_2012,ferreira_2014,costa_2014,roark_2018,lugones_2019} and the existence of a superconducting phase \citep{ruderman_1995,lander_2013,haskell_2018}.
\\
Given that the strong magnetic fields of NSs are able to deform their shape \citep{bocquet_rotating_1995,cutler_2002,oron_relativistic_2002,Dall'Osso_Shore+09a,frieben_equilibrium_2012,pili_axisymmetric_2014,gomes_2019}, and that a time-varying quadrupolar deformation leads to the emission of continuous GWs (CGWs), it is important to understand the interplay of magnetic fields and the EoS in affecting the magnetic deformation of NSs. In this sense, the existence of relations among potentially observable quantities, which are truly independent or weakly dependent on the EoS (quasi-universal relations) \citep{breu_2016,soldateschi_2021_1} may be helpful in disentangling the effects of these two major unknowns.
\\\\
In this work we apply our recent results \citep{soldateschi_2020,soldateschi_2021,soldateschi_2021_1}
regarding a quasi-universal relation linking the NS mass, radius, magnetic deformation and surface magnetic field both to the case of the Galactic pulsar population as contained in the ATNF catalogue \citep{atnf_2005} and simulated through a population synthesis approach. In particular, we assess the detectability of CGWs through the use of GW detectors, showing that a significant fraction of the MSP population in the Galaxy may be observable even with existing detectors when they reach their design sensitivity,
while canonical pulsars seem to be beyond the reach even of 3rd generation ones.

% The introduction should briefly place the study in a broad context and highlight why it is important. It should define the purpose of the work and its significance. The current state of the research field should be reviewed carefully and key publications cited. Please highlight controversial and diverging hypotheses when necessary. Finally, briefly mention the main aim of the work and highlight the principal conclusions. As far as possible, please keep the introduction comprehensible to scientists outside your particular field of research. Citing a journal paper \cite{ref-journal}. Now citing a book reference \cite{ref-book1,ref-book2} or other reference types \cite{ref-unpublish,ref-communication,ref-proceeding}. Please use the command \citep{ref-thesis,ref-url} for the following MDPI journals, which use author--date citation: Administrative Sciences, Arts, Econometrics, Economies, Genealogy, Histories, Humanities, IJFS, Journal of Intelligence, Journalism and Media, JRFM, Languages, Laws, Religions, Risks, Social Sciences.

%%%%%%%%%%%%%%%%%%%%%%%%%%%%%%%%%%%%%%%%%%
\section{Materials and Methods}
\begin{figure}[H]
\includegraphics[width=0.9\columnwidth]{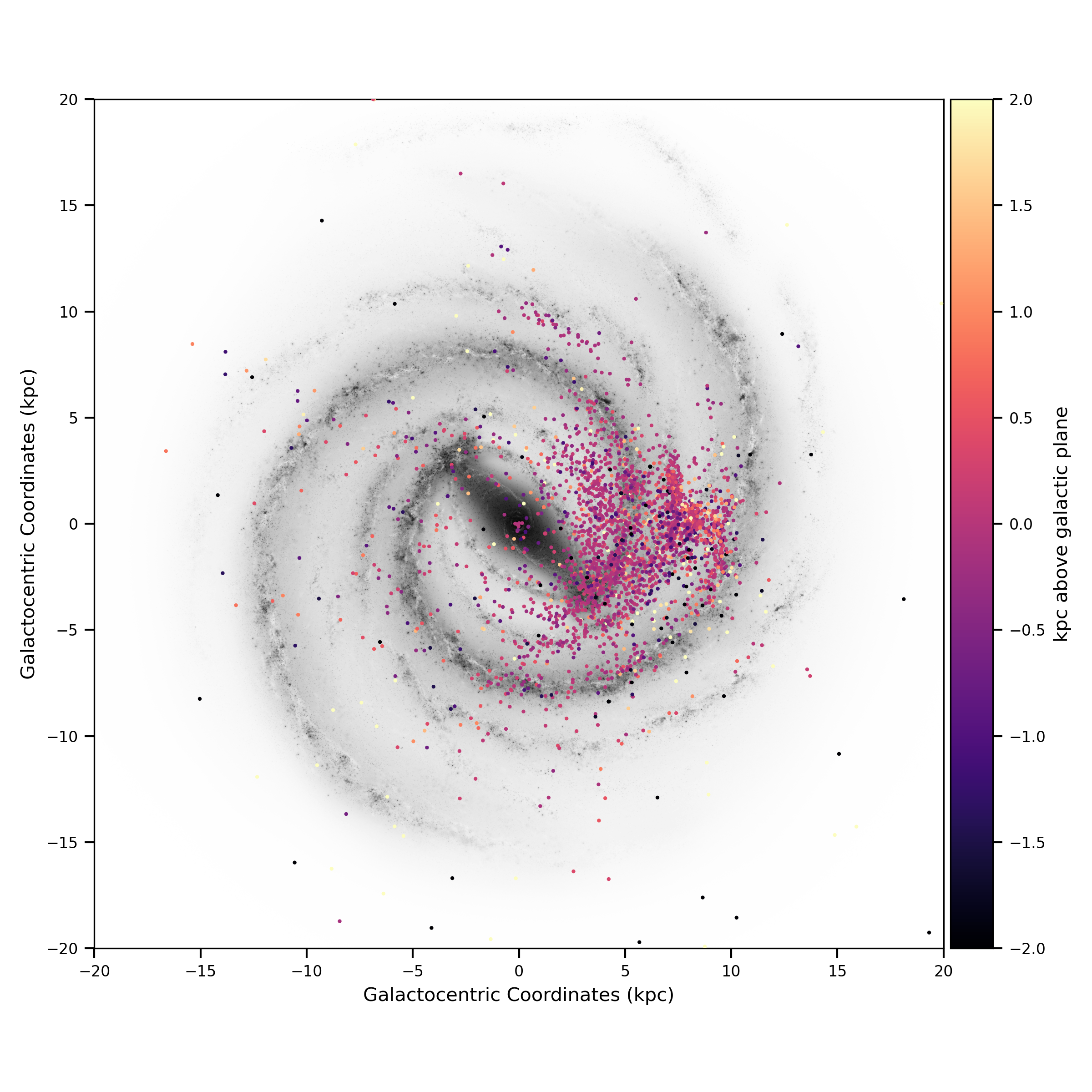}\\
\includegraphics[width=0.9\columnwidth]{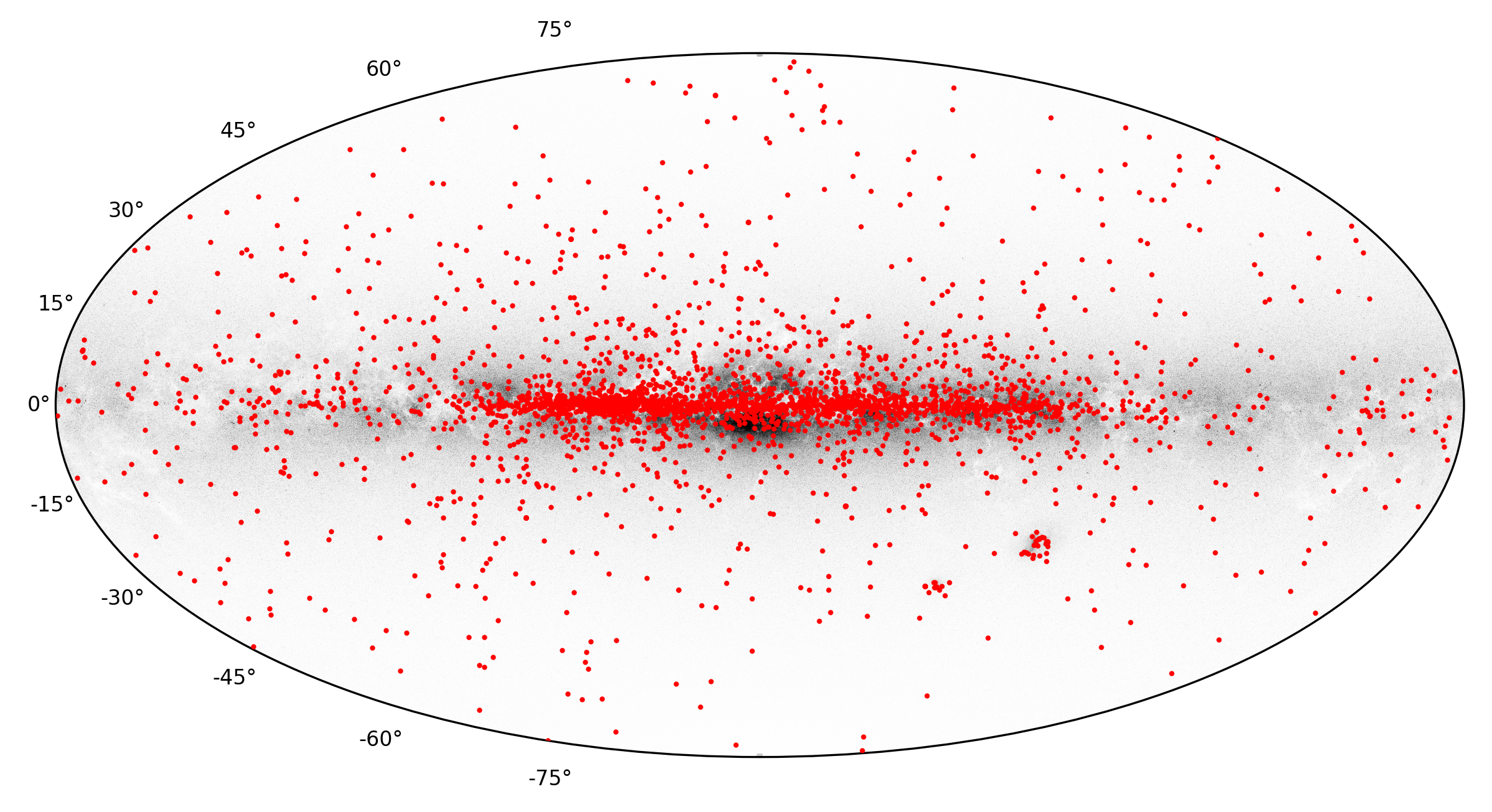}
\caption{Face-on (top plot, in galactocentric coordinates) and edge-on (bottom plot, in ICRS coordinates) rendition of the Galaxy along with the position of the NSs contained in the ATNF catalogue (case study A). The color scale in the top plot was capped at $\pm 2$ kpc for ease of visualisation. These plots were made using the \texttt{mw-plot} Python package: \url{https://pypi.org/project/mw-plot/}.\label{fig:1}}
\end{figure}

The CGWs strain $h_0$ emitted by a NS rotating with frequency $f_\mathrm{rot}$, at distance $d$ from the detector is
\begin{equation}
 h_0 = \frac{16\pi^2 G}{c^4}\frac{\mathcal{Q}f^2_\mathrm{rot}}{d}\;,
\end{equation}
where $G$ is Newton's gravitational constant, $c$ the speed of light, and $\mathcal{Q}$ is the quadrupole moment. The quadrupole moment can be written as the product of the moment of inertial $\mathcal{I}$ times the  quadrupolar deformation of the NS, $e$. In the Newtonian limit, when the deformation is caused by a purely poloidal magnetic field, the shape of the NS is axisymmetric, and one can write
\begin{equation}
  e=\bigg| \frac{I_\mathrm{zz}-I_\mathrm{xx}}{I_\mathrm{zz}} \bigg|
\end{equation}
where $I_\mathrm{xx},I_\mathrm{zz}$ are the moments of inertia of the NS, computed in the Newtonian limit, and the $z$ axis is the symmetry axis of the system. It was shown that the Newtonian value of $e$ is a good approximation for the correct GR one \citep{pili_general_2015}.
We have found \citep{soldateschi_2021,soldateschi_2021_1} that, for typical magnetic fields of NSs, the magnetic deformation $e$ of a NS is well approximated by the formula
\begin{equation}\label{eq:approx}
  e \approx c_\mathrm{s}B^2_\mathrm{s} \;,
\end{equation}
where $B_\mathrm{s}$ is the surface magnetic field at the pole, in units of $10^{18}$G, and $c_\mathrm{s}$ is called  `distortion coefficient'. Moreover, we have found, by computing $\sim 65000$ models of NSs with the \texttt{XNS} code \citep{bucciantini_general_2011,pili_axisymmetric_2014,soldateschi_2020}, that $c_\mathrm{s}$ can be approximated with great accuracy by the following quasi-universal relation:
\begin{equation}\label{eq:cs}
  c_\mathrm{s} = 2.97 R_{10}^{4.61} M_{1.6}^{-2.80} \;,
\end{equation}
where $R_{10}=R_\mathrm{c}/10$km, $M_{1.6}=M_\mathrm{k}/1.6$M$_\odot$, and $R_\mathrm{c}$ and $M_\mathrm{k}$ are the circularisation radius and the Komar mass of the NS, respectively (see \citep{gourg_2010} and \citep{gourgoulhon_3+1_2012} for their definition). This holds for several EoSs that satisfy current observational and particle physics constraints, computed according to various techniques and with different particle contents. We have found that this approximation holds also for two EoS describing strange quark stars, although with a smaller accuracy, and for this reason we do not consider those EoSs here. Moreover, it was previously found \citep{breu_2016} that also the moment of inertia $\mathcal{I}$ is well approximated by a function of just the mass and radius of the NS, for a large sample of EosS. Then, if the rotation frequency, distance, surface magnetic field, mass and radius of a NS are known, one can estimate the strain of CGWs that it should emit, independently of the EoS. However, the radii of NSs are a notoriously difficult quantity to measure, and for this reason we chose to consider the two EoSs which give the most different radii among the ones we studied (the APR4 \citep{akmal_1998,Typel_Oertel_Klaehn_2013} and the NL3$\omega \rho$ \citep{Horowitz_Piekarewicz_2001,Fortin_Providencia_Raduta_Gulminelli_Zdunik_Haensel_Bejger_2016}),
and use them to calculate the radii of the NSs from their mass. With this approach, we expect that the results obtained by considering other EoSs should be contained within the limits we find in these two cases.
\\\\
In the following we present the results obtained from two different approaches: case study A and case study B. In case A we generate a population of NSs with the following characteristics. The mass is sampled from a bimodal Gaussian distribution by \citep{antoniadis_2016}, whose peaks are located at $1.396$M$_\odot$ and $1.84$M$_\odot$. The magnetic field is sampled from a log-normal distribution \citep{faucher_2006}  with mean of $10^{12.65}$G. While this distribution is consistent with the observations of canonical pulsars contained in the ATNF catalogue, magnetic fields in MSPs are observed to have much lower values. A possible explanation for this is that the actual magnetic field of MSPs, which distorts their shape, is somehow hidden from observations, either through an accretion process \citep{bisnovatyi_1974,romani_1990} or due to ambipolar diffusion \citep{cruces_2019}. In order to avoid possible selection biases, we chose to generate also the magnetic field of canonical pulsars, even if their magnetic field have been measured. In fact, NSs in the ATNF catalogue tend to have a slightly lower magnetic field than predicted by the aforementioned distribution, possibly due to the fact that pulsars with a stronger magnetic field shut off radio emission more rapidly and have a lower chance of being detected. The rotation frequency and the distance are taken from the ATNF catalogue \citep{atnf_2005}.
This sample consists of 3177 NSs, that is the present number of NSs contained in the ATNF catalogue minus few records whose period or distance are missing.
Their position in the Galaxy can be seen in Fig.~\ref{fig:1}. Case study B consists of a generated population of $10^4$ NSs, which allows us to sample the strain distribution of pulsars with enough statistical accuracy.
The mass is computed through three possible Gaussian bimodal mass distributions: the same as in case A \citep{antoniadis_2016}; another one peaked at $1.34$M$_\odot$ and $1.78$M$_\odot$, with a maximum mass cutoff at $2.9$M$_\odot$ \citep{alsing_2018}; a third one peaked at $1.34$M$_\odot$ and $1.47$M$_\odot$\citep{farrow_2019}.
The magnetic field, as before, is sampled from a log-normal distribution \citep{faucher_2006}.
The rotation frequency is computed by fitting the frequency distribution of the ATNF pulsars and then sampling from it; the position is computed by sampling nine different possible distributions \citep{narayan_1987,lorimer_1993,lorimer_2006,kiel_2009,faucher_2010,lorimer_2012,gregoire_2013,hooper_2013,ronchi_2021}. Thus, we considered a total of 28 different populations. We note that both the surface magnetic field strength contained in the ATNF catalogue and the one sampled from the expected distribution are computed from the spin-down formula in the case of orthogonal spin and magnetic axii,
while $B_\mathrm{s}$ in Eq.~\ref{eq:approx} is that at the pole. For this reason, both the magnetic field taken from the catalogue and that sampled from the distribution need to be multiplied by a factor of 2. Moreover, we note that the strain of the plus and cross polarisations $h_{+,\times}$ of CGWs emitted at twice the rotation frequency of the NS contain a factor $\sin ^2 \alpha$, where $\alpha$ is the angle between the spin and magnetic axii \citep{bonazzola_1996}. By using the magnetic field induced deformation Eq.~\ref{eq:approx} and generating a magnetic field corresponding to that of the spin-down formula, one can obtain $h_{+,\times}$ without needing to specify the inclination angle $\alpha$, because the factor $\sin ^2 \alpha$ gets simplified.
% Materials and Methods should be described with sufficient details to allow others to replicate and build on published results. Please note that publication of your manuscript implicates that you must make all materials, data, computer code, and protocols associated with the publication available to readers. Please disclose at the submission stage any restrictions on the availability of materials or information. New methods and protocols should be described in detail while well-established methods can be briefly described and appropriately cited.
%
% Research manuscripts reporting large datasets that are deposited in a publicly avail-able database should specify where the data have been deposited and provide the relevant accession numbers. If the accession numbers have not yet been obtained at the time of submission, please state that they will be provided during review. They must be provided prior to publication.
%
% Interventionary studies involving animals or humans, and other studies require ethical approval must list the authority that provided approval and the corresponding ethical approval code.
% \begin{quote}
% This is an example of a quote.
% \end{quote}

%%%%%%%%%%%%%%%%%%%%%%%%%%%%%%%%%%%%%%%%%%
\section{Results}
\begin{figure}[H]
\includegraphics[width=\columnwidth]{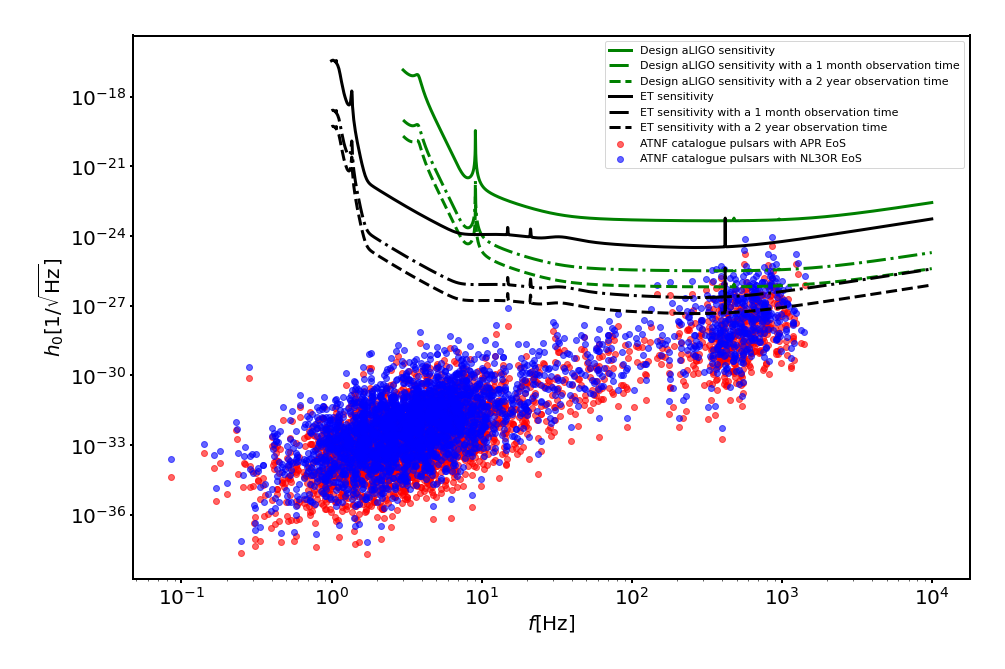}
\caption{Strain of CGWs emitted by the pulsars contained in the ATNF catalogue. Each point is a specific NS in the catalogue, and its position on the $x-$axis denotes the frequency of the emission of CGWs. The colour of the points indicates which EoS has been assumed to calculate the NS radius from its mass, either the APR (red points) or the NL3$\omega \rho$ (blue points) EoS. The solid lines are the sensitivity curves of the aLIGO (green line) and ET detectors (black line). The dot-dashed and dashed lines are the minimum detectable strain by aLIGO (green lines) and ET (black lines) in the case of a continuous 1 month (dot-dashed lines) and 2 years (dashed lines) observation time. \label{fig:2}}
\end{figure}

\begin{figure}[H]
\includegraphics[width=\columnwidth]{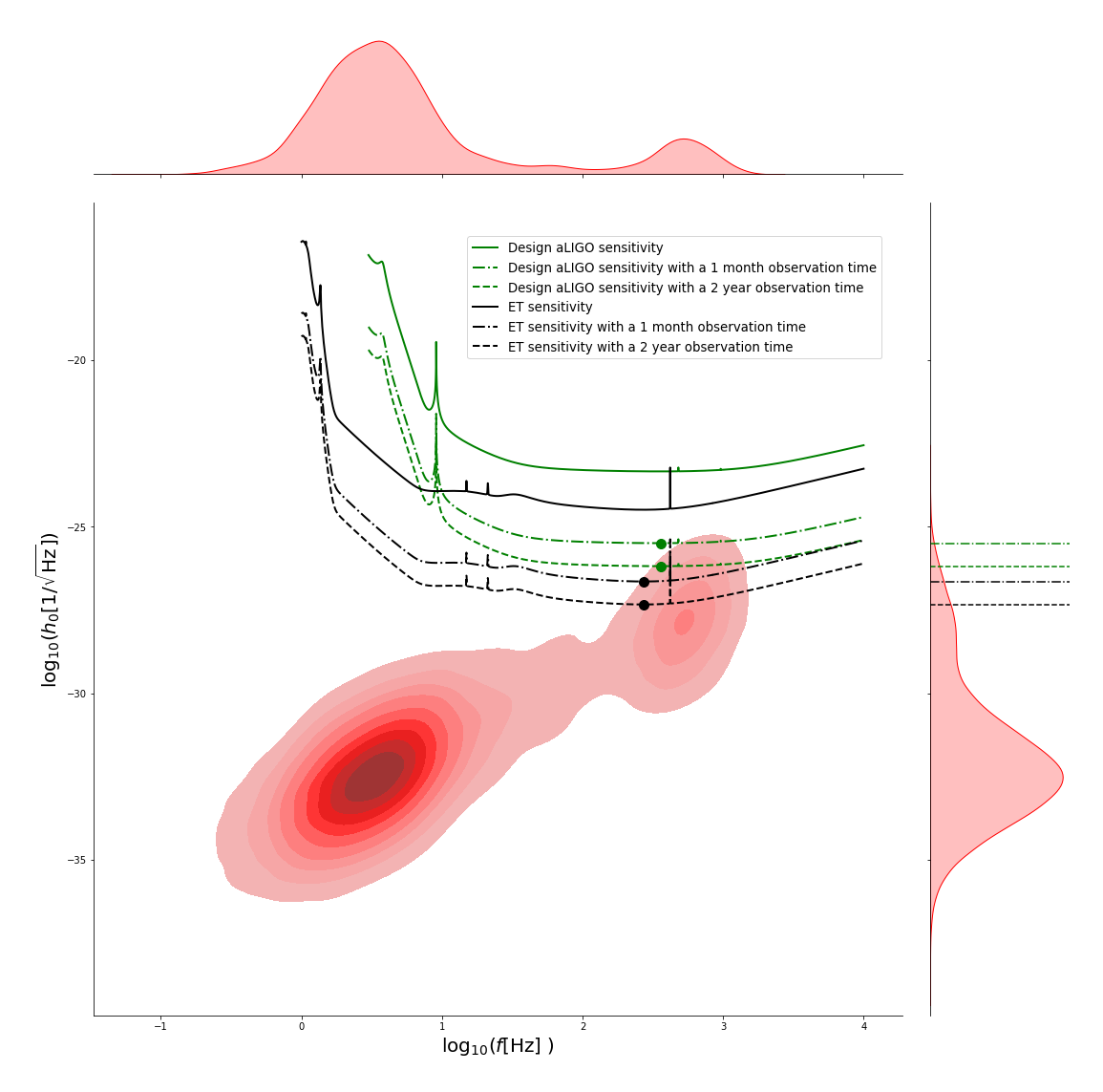}
\caption{Plot of the probability density function (red contour plot) associated with model A randomly repeated 10 times (see text for more details). The two distribution on the top and right axii are the marginal distributions. The solid lines are the sensitivity curves of the aLIGO (green line) and ET detectors (black line). The dot-dashed and dashed lines are the minimum detectable strain by aLIGO (green lines) and ET (black lines) in the case of a continuous 1 month (dot-dashed lines) and 2 years (dashed lines) observation time. The green and black points denote the minima of these curves, and the green and black lines on the axis on the right refer to the values of these minima. The fraction of the NS population that is above those lines is potentially observable with the given instrument and observing time. \label{fig:3}}
\end{figure}

In Fig.~\ref{fig:2} we can see the predicted strain of CGWs emitted by the NSs contained in the ATNF catalogue (case study A). Each point denotes a specific NS in the catalogue, its position on the $x-$axis being the frequency at which it emits CGWs, that is twice its rotation frequency. The colour of the points indicates which EoS has been assumed to calculate the NS radius from its mass, either the APR (red points) or the NL3$\omega \rho$ (blue points). The lines are the minimum detectable strain of the advanced LIGO (aLIGO) detector at design sensitivity\footnote{The aLIGO design densitivity curves can be found at \url{https://dcc.ligo.org/LIGO-T1800044/public}.} (green lines), expected to be achieved during the O4 observing run \citep{abbott_2020_sens}, and of the Einstein Telescope (ET) detector in the D configuration\footnote{The ET sensitivity curves can be found at \url{http://www.et-gw.eu/index.php/etsensitivities}.} (black lines) \citep{hild_2011_etd}. The solid lines are the nominal sensitivity curves, while the dot-dashed and dashed lines are the minimum detectable strain in the case of continuous 1 month and 2 years observation time, respectively. For a search over time $T$, the minimum detectable strain by a ground-based interferometer is \citep{watts_2008}
\begin{equation}
  h_0 \approx 11.4 \sqrt{\frac{S_\mathrm{n}}{T}}\;,
\end{equation}
where $S_\mathrm{n}$ is the power spectral density of the detector noise (thus $\sqrt{S_\mathrm{n}}$ is the nominal sensitivity curve for the detectors plotted in Fig~\ref{fig:2}.). As can be seen from Fig.~\ref{fig:2}, there are two main NSs populations contained in the ATNF catalogue, that is MSPs, emitting CGWs at a frequency $f\gtrsim 50$Hz, and canonical pulsars. We see that CGWs emitted by MSPs have a much larger strain, making them potentially observable by both aLIGO and ET with 1 month to 2 years observing time. Moreover, the only assumption regarding the EoS, that is used to compute the radii of the considered NSs, has the effect of increasing the strain by a factor of 2 to 9 when using the NL3$\omega \rho$ EoS instead of the APR EoS. On the other hand, canonical pulsars seem to be mostly invisible to even 3rd generation detectors.
\\\\
Since sampling from the expected distributions of mass and magnetic field has the effect of randomly changing the strain when generating different populations, we generated another population consisting of case A randomly repeated 10 times: for each NS in the catalogue, with its fixed rotation frequency and distance, we extracted 10 random samples from the mass and magnetic field distributions, effectively generating a population of 63540 NSs (31770 NSs for each of the two EoS). Then, we used a Gaussian kernel density estimation (KDE) procedure to estimate the probability density function of this population. The results are plotted in Fig.~\ref{fig:3}. The red contour plot is the probability density function associated to the generated NS population, while the two distribution on the top and right axii are the marginal distributions. The green and black lines are the sensitivity curves of the aLIGO and ET detectors, as in Fig.~\ref{fig:2}. The green and black points denote the minima of these curves, and the green and black lines on the axis on the right refer to the values of these minima. The fraction of the NS population that is above those lines is potentially observable with the given instrument and observing time. In particular, using the aLIGO detector with a 1 month (2 years) observation time, $\sim 3\%$ ($\sim 9\%$) of the MSP population could be detected; instead, by using the ET telescope with a 1 month (2 years) observation time, $\sim 16\%$ ($\sim 32\%$) of the MSP population could be detected. In any case, canonical pulsars seem to be out of both detectors' range. We note that these results are to be considered as the most optimistic case, as they are derived under the assumption of a purely poloidal field: if a toroidal component is present, the NSs magnetic deformation is smaller with respect to the case of a pure geometry, resulting in a lower detection rate.
Since the pulsars we studied in case A have a corresponding name and entry in the ATNF catalogue, we estimated the probability of detection of the 5 most promising ones. In order to do so, we chose the 5 NSs, described by the NL3$\omega \rho$ EoS, with the largest median strain computed by considering 100 realisations of each. Then, we computed a KDE and estimated the probability of detection by aLIGO with a 1 month and 2 years observation time. The results are reported in Tab.~\ref{tab:1}.
\begin{specialtable}[H]
\small
\caption{Top 5 pulsars in the ATNF catalogue with the highest probability of detection according to our study. The pulsar's name, distance and period are reported, as recorded in the ATNF catalogue. The median strain column reports the median value of the strain $h_0$ for each pulsar, estimated by generating 100 samples of each. The last column contains the probability of detection of each NS by aLIGO with a 1 month (2 years) observation time. \label{tab:1}}
\begin{tabular}{ccccc}
\toprule
\textbf{Name} & \textbf{Distance [kpc]}	& \textbf{Period [s]} & \textbf{Median strain $[1/\sqrt{\mathrm{Hz}}]$} & \textbf{Detection probability}\\
\midrule
J0605+3757 & 0.215 & 0.002728 & 3.21$\times 10^{-29}$ & 18\% (36\%) \\
J0636+5129 & 0.210 & 0.002869 & 3.57$\times 10^{-29}$ & 15\% (33\%) \\
J0034-0534 & 1.348 & 0.001877 & 1.57$\times 10^{-29}$ & 14\% (30\%) \\
J1400-1431 & 0.278 & 0.003084 & 1.91$\times 10^{-29}$ & 13\% (30\%) \\
J1653-0158 & 0.840 & 0.001968 & 3.58$\times 10^{-29}$ & 12\% (28\%) \\
\bottomrule
\end{tabular}
\end{specialtable}

\begin{figure}[H]
\includegraphics[width=0.49\columnwidth]{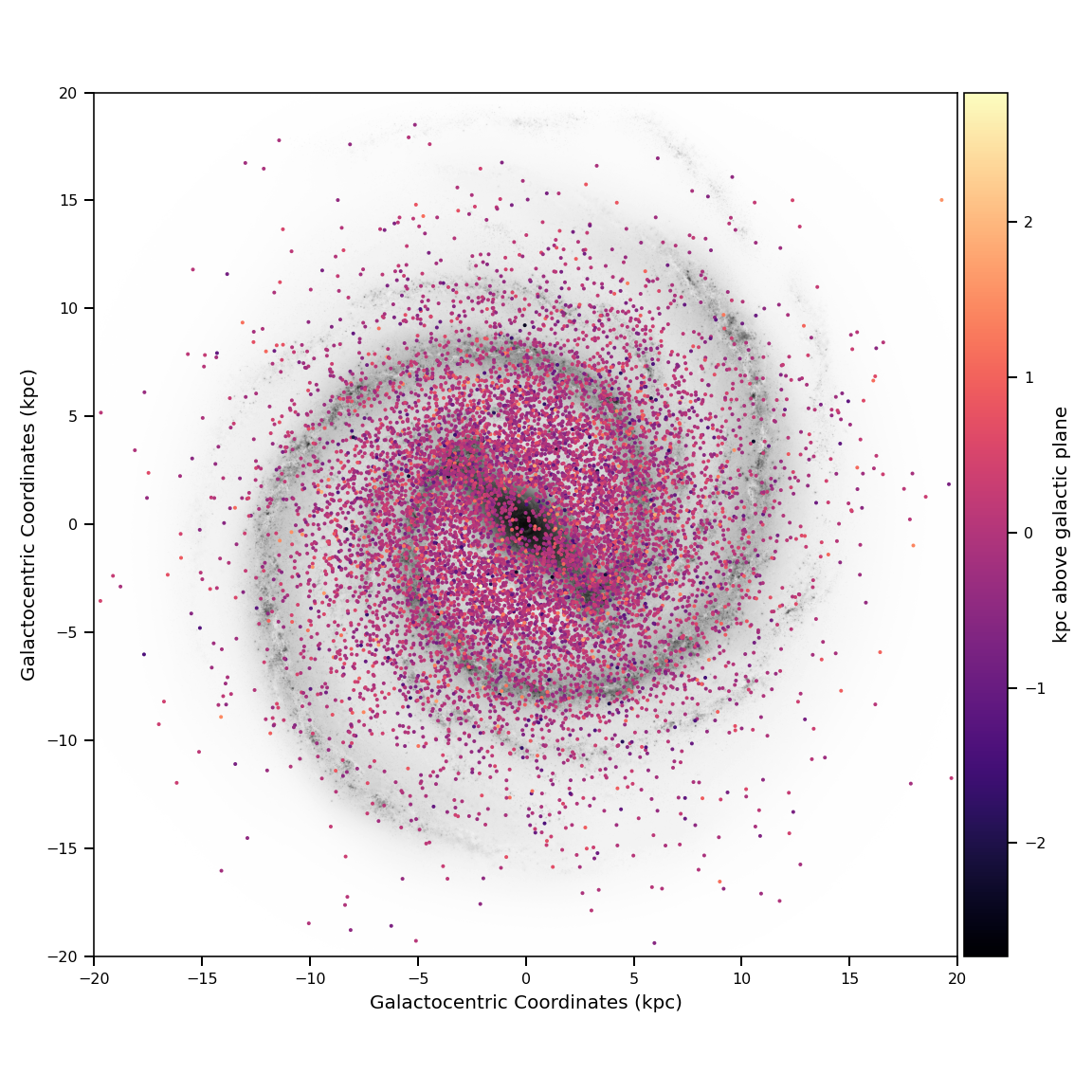}
\includegraphics[width=0.493\columnwidth]{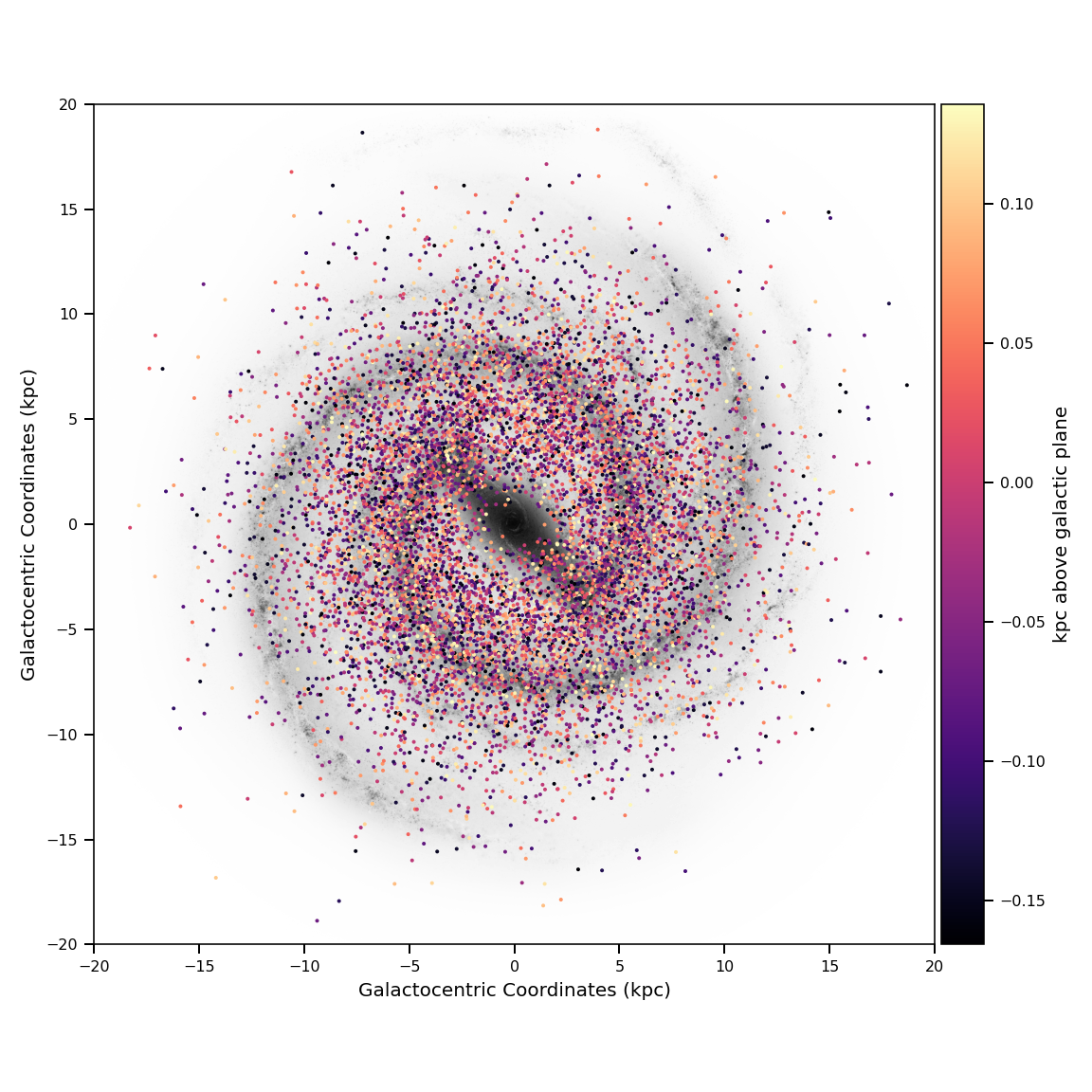}\\
\includegraphics[width=0.49\columnwidth]{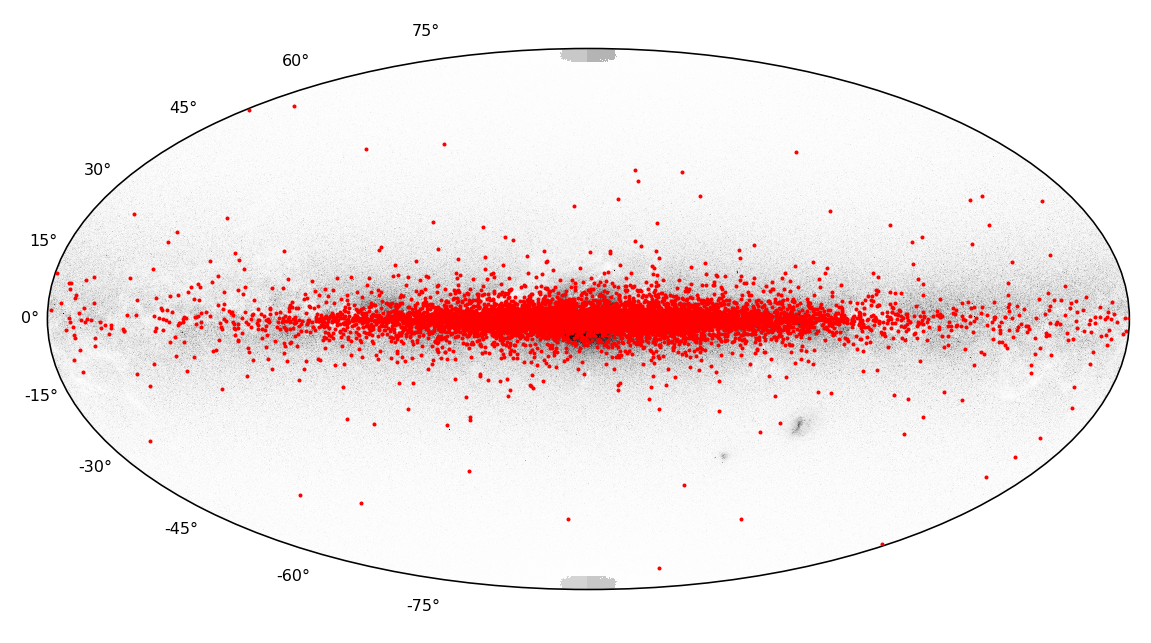}
\includegraphics[width=0.493\columnwidth]{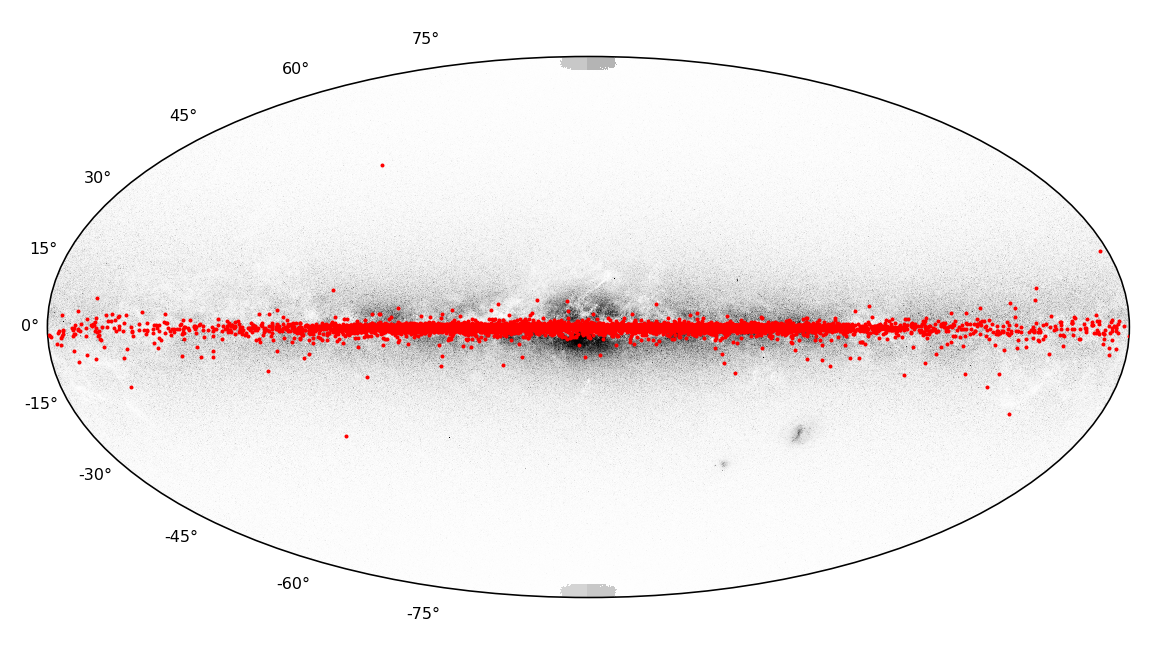}
\caption{Face-on (top plot, in galactocentric coordinates) and edge-on (bottom plot, in ICRS coordinates) rendition of the Galaxy along with the position of the NSs population generated according to the position distributions of case study B1 (left plots) and case study B2 (right plots). These plots were made using the \texttt{mw-plot} Python package: \url{https://pypi.org/project/mw-plot/}.\label{fig:4}}
\end{figure}

\begin{figure}[H]
\includegraphics[width=\columnwidth]{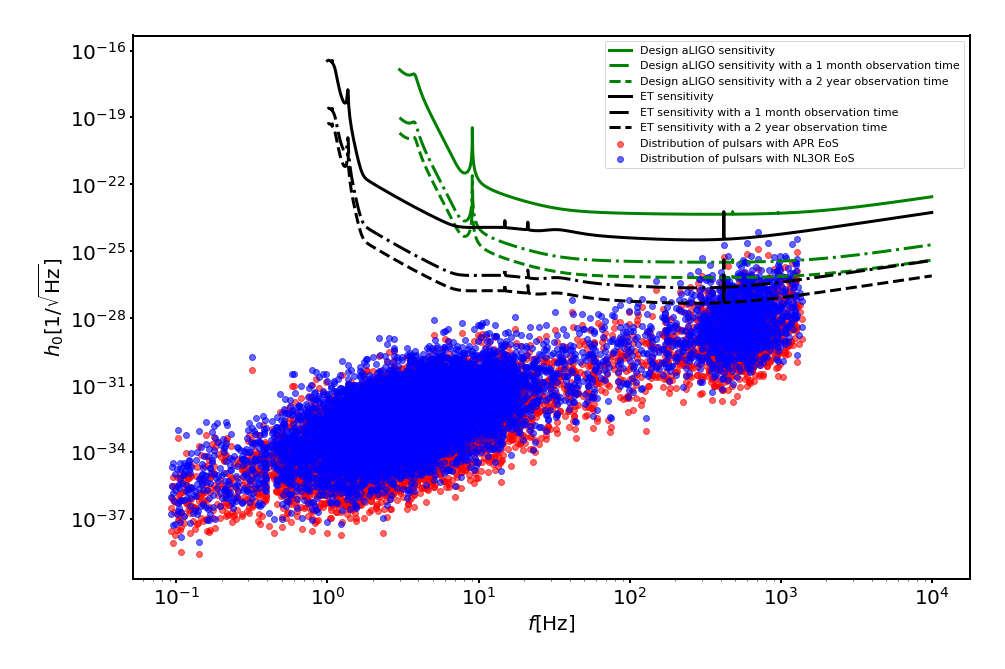}\\
\includegraphics[width=\columnwidth]{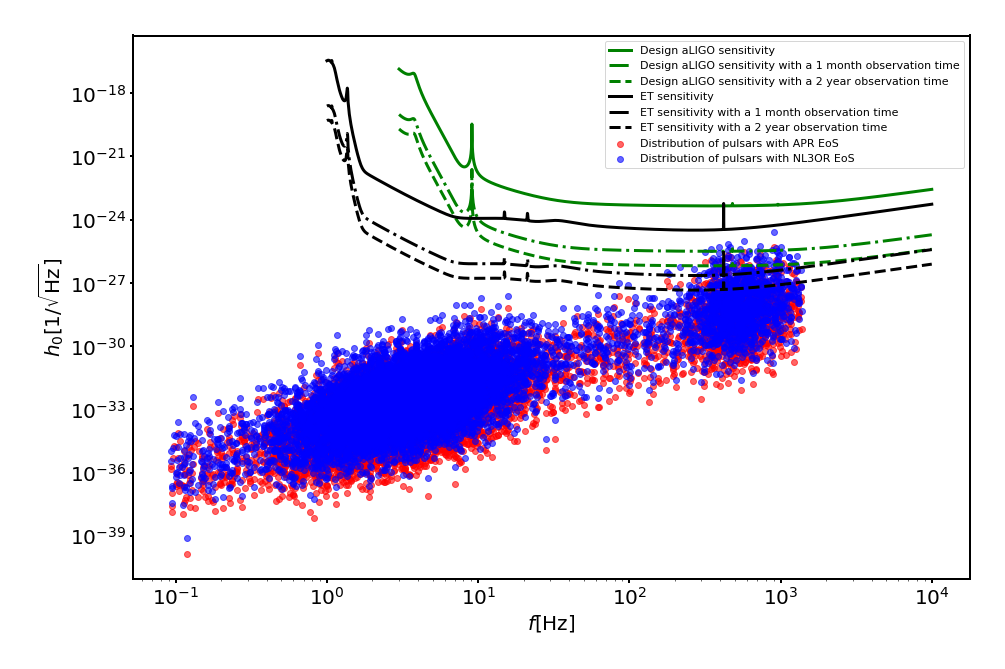}
\caption{Strain of CGWs emitted by the pulsars generated according to the models of case study B1 (top plot) and case study B2 (bottom plot). Each point is a specific NS, and its position on the $x-$axis denotes the frequency of the emission of CGWs. The colour of the points indicates which EoS has been assumed to calculate the NS radius from its mass, either the APR (red points) or the NL3$\omega \rho$ (blue points) EoS. The solid lines are the sensitivity curves of the aLIGO (green line) and ET detectors (black line). The dot-dashed and dashed lines are the minimum detectable strain by aLIGO (green lines) and ET (black lines) in the case of a continuous 1 month (dot-dashed lines) and 2 years (dashed lines) observation time. \label{fig:5}}
\end{figure}

\begin{figure}[ht]
\includegraphics[width=\columnwidth]{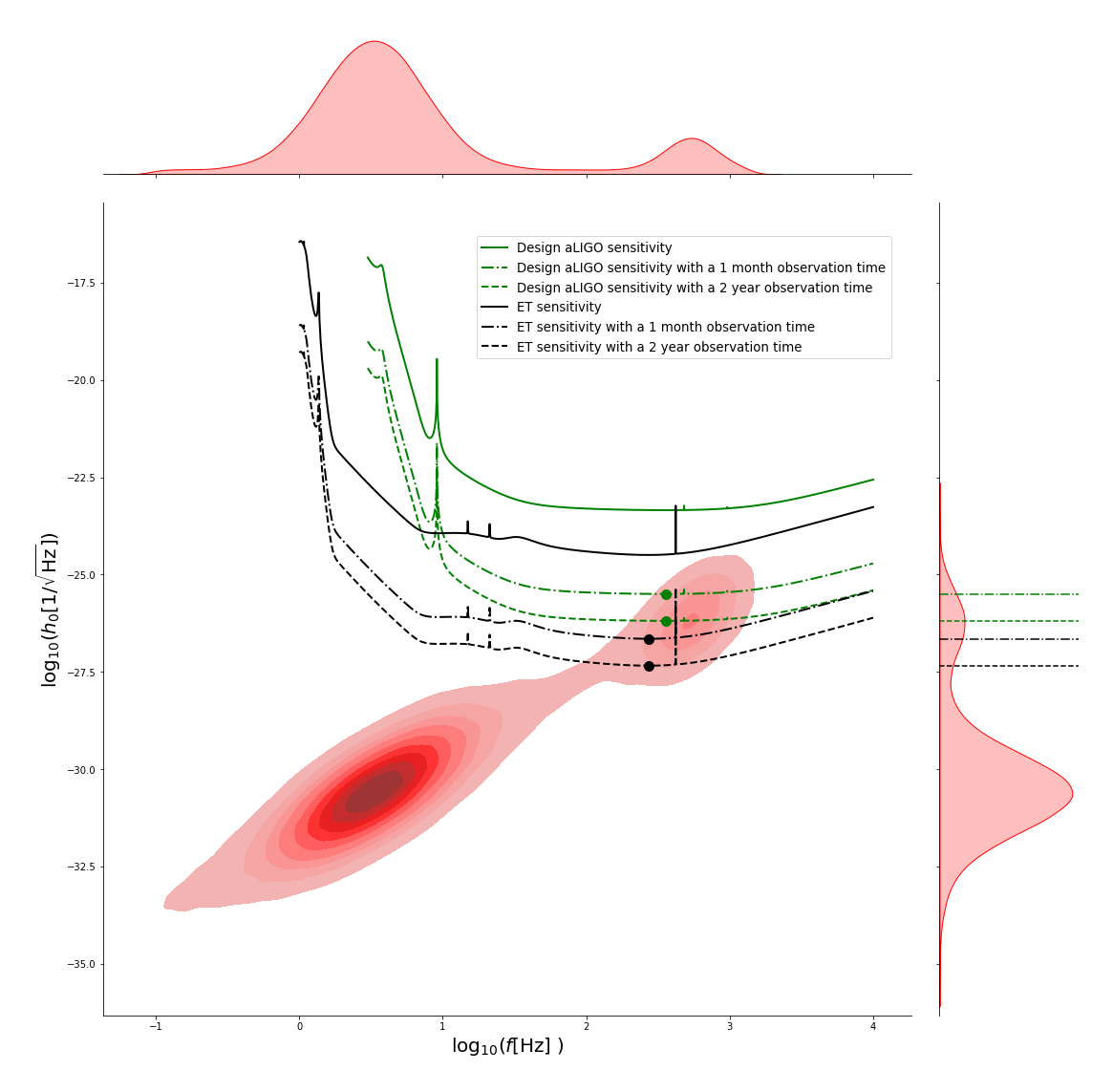}
\caption{Plot of the probability density function (red contour plot) associated with a NS population of $10^5$ samples made according to case B1 endowed with a superconducting core (see text for more details). The two distribution on the top and right axii are the marginal distributions. The solid lines are the sensitivity curves of the aLIGO (green line) and ET detectors (black line). The dot-dashed and dashed lines are the minimum detectable strain by aLIGO (green lines) and ET (black lines) in the case of a continuous 1 month (dot-dashed lines) and 2 years (dashed lines) observation time. The green and black points denote the minima of these curves, and the green and black lines on the axis on the right refer to the values of these minima. The fraction of the NS population that is above those lines is potentially observable with the given instrument and observing time. \label{fig:7}}
\end{figure}

In order to compute the strain of a more numerous samples of NSs, we consider the case study B, where we generate a population of $10^4$ NSs. This sample size allows us to sample the strain distribution of pulsars with a sufficient statistical accuracy. We found that all the combinations of mass and position distributions give similar results regarding the strain distribution. For this reason, in the following we show only the combinations of one mass distribution (the same as in case A \citep{antoniadis_2016}) and two position distributions \citep{lorimer_2006,kiel_2009}, denoted in the following as case B1 and case B2, respectively. In the first case the radial distribution of NSs on the Galactic plane is given by a gamma distribution peaked at $\sim 5.0$ kpc from the Galactic centre (model C in the paper \citep{lorimer_2006}), while the distribution of their height with respect to the Galactic plane is given by an exponential distribution with a scale height of $330$ pc (model S in the paper). In the second case, the radial distribution is that of NSs at birth \citep{yusifov_2004}, shaped as a gamma distribution peaked at $\sim 6.2$ kpc from the Galactic centre, while the height at birth is given by a uniform distribution between $150$ pc and $-150$ pc (model C' in the paper \citep{kiel_2009}).
We note that we considered two very different NSs populations - an evolved population (case B1) and a population at birth (case B2) - because our results show that their position in the Galaxy does not seem to significantly influence their visibility by CGWs. Moreover, we note that the biggest difference between cases B1 and B2 lies in the distribution of heights above the Galactic plane, which has a scale length that is much lower than that of the radial distribution. For this reason, it is expected that these two cases give similar results.
The position of the pulsar population generated according to case B1 and case B2 are shown in Fig.~\ref{fig:4} on the left and right, respectively.
In fig.~\ref{fig:5} we plot the resulting strain distributions for case study B1 and B2 (top and bottom plots, respectively).
We clearly see that the differences in the the resulting strain distribution are minimal, even though the positions of the two populations have a susbstantially different shape (see Fig.~\ref{fig:4}). In order to estimate the probability density distribution through KDE we increased the number of samples to generate to $10^5$ for each EoS, resulting in a total population of $2\times 10^5$ NSs. Given that cases B1 and B2 give practically equivalent results regarding $h_0$, we only plot the density obtained from case B1 in Fig.~\ref{fig:6}. We see that, using the aLIGO detector with a 1 month (2 years) observation time, $\sim 1\%$ ($\sim 5\%$) of the MSP population could be detected; instead, by using the ET telescope with a 1 month (2 years) observation time, $\sim 10\%$ ($\sim 23\%$) of the MSP population could be detected. As we found for case A, canonical pulsars seem to be out of both detectors' range.
\\
In the case of a NS endowed with a superconducting core, the extent to which the magnetic field can deform the NS is much more enhanced \citep{cutler_2002,frieben_equilibrium_2012}. In this case, we expect NS models to develop a distortion coefficient that is roughly $B_\mathrm{c1}/\langle B \rangle$ times higher than without a superconducting core, where $B_\mathrm{c1} \approx 10^{15}$G is the first critical magnetic field strength and $\langle B \rangle$ is the volume average of the magnitude of the magnetic field $B$ \citep{soldateschi_2021_1}. As we show in Fig.~\ref{fig:7}, the fraction of observable CGWs emitted by MSPs is greatly increased in this case: $\sim 18\%$ ($\sim 48\%$) using the aLIGO detector with a 1 month (2 years) observation time and $\sim 69\%$ ($\sim 90\%$) using the ET telescope with a 1 month (2 years) observation time. While the strain of CGWs emitted by canonical pulsars is certainly enhanced by the presence of a superconducting core, next generation telescopes like ET still fall short of the required sensitivity of at least one order of magnitude.

%%%%%%%%%%%%%%%%%%%%%%%%%%%%%%%%%%%%%%%%%%
\section{Discussion}
We have shown that the quasi-universal relation Eq.~\ref{eq:cs} linking the magnetic deformation of a NS to its mass, radius and surface magnetic field can be used to compute the strain of the CGWs they emit in a way that is independent of their EoS. This can be done once the NS mass, radius, surface magnetic field, rotation period and distance are known.
Measuring directly the radius is notoriously difficult; however, once an EoS is assumed, there is a one to one relation with the mass, a much easier quantity to estimate, even from a statistical point of view.
\begin{figure}[H]
\includegraphics[width=\columnwidth]{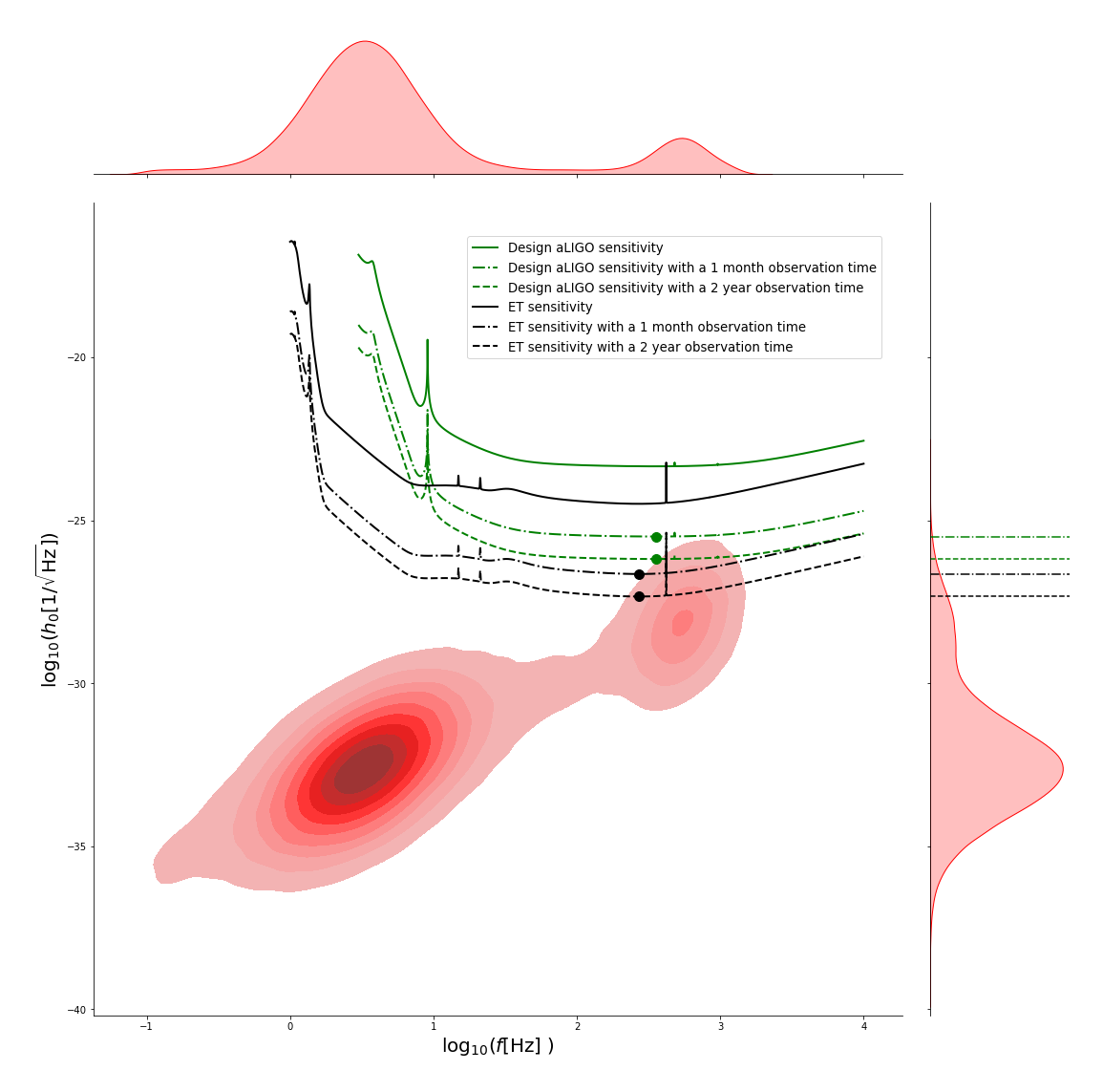}
\caption{Plot of the probability density function (red contour plot) associated with a NS population of $10^5$ samples made according to case B1 (see text for more details). The two distribution on the top and right axii are the marginal distributions. The solid lines are the sensitivity curves of the aLIGO (green line) and ET detectors (black line). The dot-dashed and dashed lines are the minimum detectable strain by aLIGO (green lines) and ET (black lines) in the case of a continuous 1 month (dot-dashed lines) and 2 years (dashed lines) observation time. The green and black points denote the minima of these curves, and the green and black lines on the axis on the right refer to the values of these minima. The fraction of the NS population that is above those lines is potentially observable with the given instrument and observing time. \label{fig:6}}
\end{figure}
For this reason, we have chosen the two EoSs that give the most different radii, for the same NS mass, among the ones we used to infer the quasi-universal relation for $c_\mathrm{s}$: this way, we expect that our results regarding the detectability of the Galactic NS population should encompass a much larger selection of possible NS EoS. As we discussed, the strain computed by adopting these two EoS can differ by up to an order of magnitude. Regarding the other quantities, we adopted two different approaches. In case A we used the values for the rotation period and distance of the known pulsars in the Galaxy contained in the ATNF catalogue, and we extracted their mass and surface magnetic field from the expected distributions. In the other case B, we synthesised the whole Galactic NS population by extracting all the necessary quantities from the expected distributions, allowing us to compute the strain of potentially undetected sources.
In the last case, we chose a variety of distributions for the mass and distance of the expected NS population in the Galaxy, finding very similar results in all cases. Then, for each EoS we randomly generated a large population, and estimated the probability density distribution of $h_0$ using a KDE approach. This allows us to estimate the fraction of NSs in the Galaxy whose CGWs are within the range of ground-based future GW detectors. In particular, we considered the cases of the aLIGO and the ET detectors, both in the case of a continuous 1 month and 2 years observation time. In case A (see Fig.~\ref{fig:3}) we found that up to $\sim 9\%$ and $\sim 32\%$ of the total MSPs population lies within the reach of aLIGO and ET, respectively, considering a 2 years observing time. This amounts to a number of $\sim 270$ and $\sim 960$ detectable pulsars if one considers the expected number $\sim 3\times 10^{3}$ of MSPs within $5$ kpc of the Sun \citep{lorimer_2008}.
We note that those are the NSs such that the radio beaming intercepts our line of sight, and as such there exist a fraction of the total pulsar population which the ATNF catalogue does not account for but which may be observable through their CGWs emission.
Lower fractions are obtained in case B1 (see Fig.~\ref{fig:6}): up to $\sim 5\%$ and $\sim 23\%$ of the total MSP population with aLIGO and ET, respectively, corresponding to $\sim 2000$ and $\sim 9200$ NSs considering the $\sim 4\times 10^{4}$ MSPs expected to be present in the Galaxy \citep{lorimer_2008}.
We believe that this is due to a selection bias: as we see in Fig.~\ref{fig:1} - top plot - the NS population that is observed is, as expected, roughly centred on the position of the Solar System, which lies at Galactocentric coordinates $(x,y,z)=(8.122,0,0.021)$ kpc, and many more pulsars close to the Sun have been observed than those further away. On the other hand, as we see in Fig.~\ref{fig:4}, the expected distributions of the NSs positions are computed for  all NSs in the Galaxy, thus lowering the fraction of detectable NSs.
Finally, we explored the case of NSs endowed with a superconducting core. In this case, the effective magnetic field that deforms the star is much stronger than without the effect of superconductivity, resulting in a greater emission of CGWs by the same NSs. In fact, in this case the fraction of detectable MSPs can reach values up to $\sim 48\%$ and even $\sim 90\%$ in the case of aLIGO and ET, respectively, for an observation of 2 years, corresponding to $\sim 19200$ and $\sim 36000$ NSs. Even with just one month of observing time, $\sim 18\%$ and $\sim 69\%$ of the MSPs in the Galaxy lie within the reach of aLIGO and ET, amounting to $\sim 7200$ and $\sim 27600$ NSs respectively. In all cases we considered, CGWs emitted by canonical pulsars seem to be far too weak even for 3rd generation ground-based GW detectors, due to their slow rotation period. On the one hand, given that such a large fraction of MSPs could be detectable by aLIGO and ET in the case of superconductivity, the absence of any CGWs detection could itself be an indication of the lack of a superconducting core, effectively constraining the possibility of its existence. On the other hand, we note that all our results assume that the magnetic field inside NSs has a purely poloidal geometry. While this is clearly a simplifying assumption, it allows one to possibily infer information on the geometry of the internal magnetic field of MSPs: the presence of a toroidal component counteracts the effect of the poloidal component, effectively reducing the deformation with respect to NSs endowed with a purely poloidal field and potentially rendering invisible MSPs which, given their characteristics, should be detectable by GW detectors according to our study.
Finally, we note that we focused on the strain $h_0$ as a way to measure the detectability of NSs independently from their orientation with respect to the detectors and the inclination between the magnetic and the rotation axii. Of course, the magnetic deformation of a NS does not lead to the emission of CGWs if the magnetic axis is aligned with the rotation axis. Moerover, GW detectors have a particular antenna pattern which renders them more or less sensitive to waves coming from certain angular positions in the sky. While we believe our work can give a comprehensive overview of what to expect in terms of CGWs emission by pulsars in the Galaxy, a more in depth extension would be to consider also the expected distribution of the relative inclination between the two axii and to consider the time-varying angular position of the NSs systems with respect to the ground-based detectors on Earth.

% Authors should discuss the results and how they can be interpreted from the perspective of previous studies and of the working hypotheses. The findings and their implications should be discussed in the broadest context possible. Future research directions may also be highlighted.

%%%%%%%%%%%%%%%%%%%%%%%%%%%%%%%%%%%%%%%%%%
%\section{Conclusions}

% This section is not mandatory, but can be added to the manuscript if the discussion is unusually long or complex.

%%%%%%%%%%%%%%%%%%%%%%%%%%%%%%%%%%%%%%%%%%
% \section{Patents}

% This section is not mandatory, but may be added if there are patents resulting from the work reported in this manuscript.

%%%%%%%%%%%%%%%%%%%%%%%%%%%%%%%%%%%%%%%%%%
\vspace{6pt}

%%%%%%%%%%%%%%%%%%%%%%%%%%%%%%%%%%%%%%%%%%
%% optional
%\supplementary{The following are available online at \linksupplementary{s1}, Figure S1: title, Table S1: title, Video S1: title.}


%%%%%%%%%%%%%%%%%%%%%%%%%%%%%%%%%%%%%%%%%%
\authorcontributions{Conceptualization, Jacopo Soldateschi and Niccolò Bucciantini; Data curation, Jacopo Soldateschi and Niccolò Bucciantini; Formal analysis, Jacopo Soldateschi and Niccolò Bucciantini; Investigation, Jacopo Soldateschi and Niccolò Bucciantini; Writing – original draft, Jacopo Soldateschi; Writing – review \& editing, Niccolò Bucciantini. All authors have read and agreed to the published version of the manuscript.}

\funding{The authors acknowledge financial support from the Accordo Attuativo ASI-INAF n. 2017-14-H.0 `On the escape of cosmic rays and
their impact on the background plasma', the PRIN-INAF 2019 `Short gammaray burst jets from binary neutron star mergers' and from the INFN Teongrav collaboration.}

\conflictsofinterest{The authors declare no conflict of interest.}

\end{paracol}
%%%%%%%%%%%%%%%%%%%%%%%%%%%%%%%%%%%%%%%%%%
% To add notes in main text, please use \endnote{} and un-comment the codes below.
%\begin{adjustwidth}{-5.0cm}{0cm}
%\printendnotes[custom]
%\end{adjustwidth}
%%%%%%%%%%%%%%%%%%%%%%%%%%%%%%%%%%%%%%%%%%
\reftitle{References}


\externalbibliography{yes}
\bibliography{galaxies.bib}

% The following MDPI journals use author-date citation: Admsci,  Arts, Econometrics, Economies, Genealogy, Humanities, IJFS, Jintelligence, JRFM, Languages, Laws, Literature, Religions, Risks, Social Sciences. For those journals, please follow the formatting guidelines on http://www.mdpi.com/authors/references
% To cite two works by the same author: \citeauthor{ref-journal-1a} (\citeyear{ref-journal-1a}, \citeyear{ref-journal-1b}). This produces: Whittaker (1967, 1975)
% To cite two works by the same author with specific pages: \citeauthor{ref-journal-3a} (\citeyear{ref-journal-3a}, p. 328; \citeyear{ref-journal-3b}, p.475). This produces: Wong (1999, p. 328; 2000, p. 475)

%%%%%%%%%%%%%%%%%%%%%%%%%%%%%%%%%%%%%%%%%%
%% for journal Sci
%\reviewreports{\\
%Reviewer 1 comments and authors’ response\\
%Reviewer 2 comments and authors’ response\\
%Reviewer 3 comments and authors’ response
%}
%%%%%%%%%%%%%%%%%%%%%%%%%%%%%%%%%%%%%%%%%%
\end{document}